\documentclass[pra,nofootinbib,preprintnumbers,amssymb,amsfonts,amsmath,superscriptaddress,showpacs,hyperref]{revtex4-1}

\usepackage{csquotes}
\usepackage{epsf,epsfig,amsmath,amssymb,dsfont}
\usepackage{amsthm,mathrsfs} 
\usepackage{graphicx} 
\usepackage{float}
\usepackage{slashed}
\usepackage{float}
\usepackage{hyperref} 
\usepackage[]{subfigure}
\usepackage{multirow}
\usepackage{mathtools}
\usepackage{color}
\usepackage{cancel}
\usepackage[framemethod=TikZ]{mdframed}
\mdfdefinestyle{MyFrame}{%
    linecolor=blue,
    outerlinewidth=2pt,
    roundcorner=20pt,
    innertopmargin=\baselineskip,
    innerbottommargin=\baselineskip,
    innerrightmargin=20pt,
    innerleftmargin=20pt,
    backgroundcolor=gray!50!white}

\newcommand{\be}{\begin{equation}}
\newcommand{\ee}{\end{equation}}
\newcommand{\baln}{\begin{align}}
\newcommand{\ealn}{\end{align}}
\newcommand{\ben}{\begin{equation*}}
\newcommand{\een}{\end{equation*}}

\long\def\symbolfootnote[#1]#2{\begingroup%
\def\thefootnote{\fnsymbol{footnote}}\footnote[#1]{#2}\endgroup}

\usepackage{tikz}
\usetikzlibrary{calc}

\xdefinecolor{mycolor}{RGB}{62,96,111} 
\colorlet{bancolor}{mycolor}

\begin{document}

\title{Information Content of the Gravitational Field of a Quantum Superposition}

\author{Alessio Belenchia} 
\email{Corresponding author: a.belenchia@qub.ac.uk}
\affiliation{Centre for Theoretical Atomic, Molecular, and Optical Physics, School of Mathematics and Physics, Queen's University, Belfast BT7 1NN, United Kingdom.}

\author{Robert M. Wald}
\email{rmwa@uchicago.edu} 
\affiliation{Enrico Fermi Institute and Department of Physics, The University of Chicago, 5640 South  Ellis Avenue,  Chicago,  Illinois 60637,  USA}

\author{Flaminia Giacomini} 
\email{flaminia.giacomini@univie.ac.at }
\affiliation{Institute for Quantum Optics and Quantum Information (IQOQI), Boltzmanngasse 3 1090 Vienna, Austria.}

\author{Esteban Castro-Ruiz} 
\email{esteban.castro.ruiz@univie.ac.at }
\affiliation{Institute for Quantum Optics and Quantum Information (IQOQI), Boltzmanngasse 3 1090 Vienna, Austria.}

\author{\v{C}aslav Brukner} 
\email{caslav.brukner@univie.ac.at }
\affiliation{Institute for Quantum Optics and Quantum Information (IQOQI), Boltzmanngasse 3 1090 Vienna, Austria.}

\author{Markus Aspelmeyer} 
\email{markus.aspelmeyer@univie.ac.at }
\affiliation{Institute for Quantum Optics and Quantum Information (IQOQI), Boltzmanngasse 3 1090 Vienna, Austria.}

\date{\today}

\begin{abstract}
When a massive quantum body is put into a spatial superposition, it is of interest to consider the quantum aspects of the gravitational field sourced by the body. We argue that in order to understand how the body may become entangled with other massive bodies via gravitational interactions, it must be thought of as being entangled with its own Newtonian-like gravitational field. Thus, a Newtonian-like gravitational field must be capable of carrying quantum information. Our analysis supports the view that table-top experiments testing entanglement of systems interacting via gravity do probe the quantum nature of gravity, even if no ``gravitons'' are emitted during the experiment.
\\
\\
*Corresponding author
\\
\\
\textit{First prize essay written for the Gravity Research Foundation 2019 Essays on Gravitation}

\end{abstract}
\maketitle


General Relativity and Quantum Mechanics are pillars of modern theoretical physics and are well tested in a wide variety of regimes. Nevertheless, we still lack a satisfactory phenomenological basis for a full theory of quantum gravity. Thus, it is of considerable interest to analyze situations where both quantum theory and gravity play an essential role, as already suggested by Feynman in the 1950s~\cite{cecile2011role}. Since then, many Gedankenexperimente and, more recently, actual experimental proposals have been put forward~\cite{baym2009two,Mari:2015qva,PhysRevLett.119.240401,PhysRevLett.119.240402,carlesso2017cavendish,PhysRevA.71.024101,Anastopoulos:2015zta}.

In this essay, we make use of one of these Gedankenexperimente---originally proposed in~\cite{Mari:2015qva} and previously analyzed by us in~\cite{PhysRevD.98.126009}---to gain insight into ``where and how'' quantum information about a quantum superposition of a source particle is stored in its gravitational field, and under what circumstances it can be accessed. We will argue here that a quantum massive particle should be thought of as being entangled with its own Newtonian-like gravitational field, and thus that a Newtonian-like gravitational field can transmit quantum information.

Let us start by introducing the Gedankenexperiment, which has both a gravitational and electromagnetic version. We have two parties, Alice and Bob, at a distance $R$ from each other, each controlling a charged/massive body, with charges $q_{A}$ and $q_{B}$ and masses $m_{A}$ and $m_{B}$, respectively. We assume that Alice's particle also has spin and that, in the distant past, she sent her particle through a Stern-Gerlach apparatus, putting it in an equal superposition, $(|L \rangle_A|\downarrow\rangle_A + | R \rangle_A|\uparrow\rangle_A )/\sqrt{2}$, of states $| L \rangle_A $ and $| R \rangle_A $ of its center of mass (CM), spatially separated by distance $d\ll R$. We assume this process took place adiabatically, so that negligible (electromagnetic or gravitational) radiation was emitted.  Bob's particle is initially held in a trap with a sufficiently strong confining potential so that any influences of Alice's particle on the state of Bob's particle are negligible. 

At a pre-arranged time, $t=0$, Bob makes a choice of either releasing his particle from the trap or leaving it in the trap. If he releases his particle, then his particle will react to the electromagnetic or gravitational influence of Alice's particle, corresponding to the states $|L\rangle_A,\, |R\rangle_A$. In this case, we denote $\delta x$ the center of mass displacement of the different possible locations of Bob's particle at time $T_B$, when he completes his experiment. If $\delta x$ is sufficiently large, the location difference will make the possible states of Bob's particle nearly orthogonal, so his particle will be nearly maximally correlated with Alice's, and thus Alice's particle will be in a highly mixed state. In other words, Bob has acquired maximal ``which-path'' information about Alice's particle.

\begin{figure}[h!]
\centering
\includegraphics[width=0.45\textwidth]{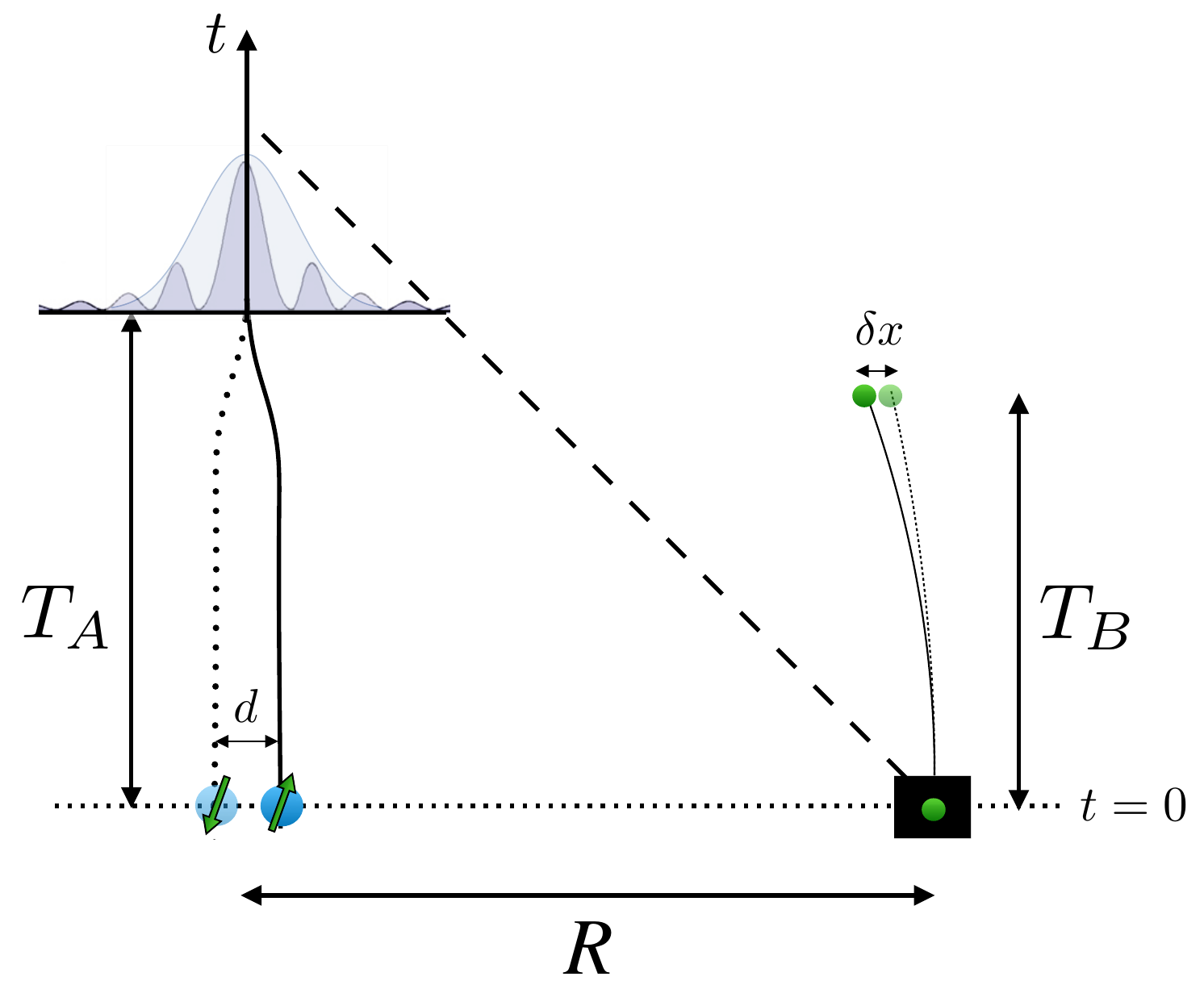}
 \caption{Arrangement of the Gedankenexperiment. Alice's particle (left) is prepared in a spatial superposition with separation $d$ while Bob's particle (right), at distance $R\gg d$, is initially localized by a trap. At the start of the protocol Bob can decide whether or not to release his particle from the trap, while Alice starts to recombine the paths of her particle. (When dividing and recombining the paths of her particle, Alice uses Stern-Gerlach devices, as discussed in~\cite{Mari:2015qva,PhysRevLett.119.240401}, for which the spin internal degree of freedom is instrumental.)}
 \label{protocol}
\end{figure}

At the same time,  Alice sends her particle through a ``reversing'' Stern-Gerlach apparatus, in such a way that if her particle had remained unentangled (and thus in a pure state), she could successfully perform an interference experiment. She completes this process in time $T_A$. It should be noted here that Alice's internal spin degree of freedom allows us to consider the case in which, if Bob is able to acquire which-path information, the entanglement at the end of the experiment is between the position of Bob's particle and the spin degree of freedom of Alice's particle, thereby greatly simplifying quantum interference tests. The arrangement of this Gedankenexperiment is illustrated in Fig.~\ref{protocol}.

This Gedankenexperiment appears to yield paradoxical results when\footnote{Here and in the following we work with units $c=\hbar=1$.} $T_A,T_B < R$. In that case, Bob's actions cannot causally influence Alice's particle, so one might expect that Alice's particle should remain in a coherent superposition. But Bob should be able to obtain ``which path'' information in his experiment, implying that his particle must be entangled with Alice's. Therefore Alice's particle cannot be in a coherent superposition. This apparent paradox was resolved in~\cite{PhysRevD.98.126009}. It was shown there that in the electromagnetic case, if the effective electric dipole ${\mathcal D}_A$ resulting from the path separation of Alice's particle satisfies 
\begin{equation}\label{dipoleem}       
{\mathcal D}_A < T_A<R,
\end{equation}
then vacuum fluctuations prevent Bob from acquiring ``which path'' information in time $T_B < R$, and his particle will not become entangled with Alice's. On the other hand, when ${\mathcal D}_A > T_A$ Alice's particle will be entangled with electromagnetic radiation emitted during the ``recombination'' of her particle, and her particle will not be in a coherent superposition, independently of what Bob does. In this case, there is no contradiction with Bob also obtaining ``which path'' information. Similarly, in the gravitational case, if the effective gravitational quadrupole\footnote{It should be noted that, it is momentum conservation that implies the  absence of dipole gravitational radiation.} ${\mathcal Q}_A $ resulting from the path separation satisfies
\begin{equation} \label{qurad}
{\mathcal Q}_A < T^2_A<R^2 
\end{equation}
then vacuum fluctuations prevent Bob from acquiring ``which path'' information, whereas if ${\mathcal Q}_A > T^2_A$, Alice's particle becomes entangled with gravitational radiation.

Although the above analysis of~\cite{PhysRevD.98.126009} resolved the apparent paradox when $T_A,T_B < R$, there are further implications that we wish to explore here. These are most easily seen if, for the electromagnetic version, we consider the case where ${\mathcal D}_A > R$ and again require $T_B < R$, but now we take $T_A \gg {\mathcal D}_A$. Similarly, for the gravitational version, we consider the case where ${\mathcal Q}_A > R^2$, $T_B < R$, and $T^2_A \gg {\mathcal Q}_A$. In this situation, Bob can acquire ``which path'' information in time $T_B < R$, but Alice has time to do the recombination adiabatically, so that negligible electromagnetic/gravitational radiation is emitted by Alice's particle. Since Bob can acquire ``which path'' information, Alice's particle must be entangled with Bob's at the end of the process. 

Since Alice's particle is entangled with Bob's, it cannot be in a coherent superposition at the end of the process.
There is no causality paradox in this case, since we necessarily have $T_A > R$, so the state of Alice's particle can be causally influenced by what Bob's particle does. Nevertheless, since $T_B < R$, Bob's particle cannot be influenced by Alice's actions during the time of his experiment. For all that Bob knows, Alice may have recombined her superposition in time $T_A < R$. 

If Alice had actually recombined her superposition in time $T_A < R$, but still $D_A>R$, we would be in the original case discussed above, where Alice's particle emits entangling radiation during the recombination. The final state of Alice, Bob, and the radiation field would be
\begin{equation}
\sim|x=0\rangle_{A}\otimes(|\uparrow\rangle_{A}|\phi_R\rangle_{\rm{rad}}|L\rangle_B+|\downarrow\rangle_{A}|\phi_L\rangle_{\rm{rad}}|R\rangle_B),
\end{equation}
where $|\phi_{R(L)}\rangle_{\rm{rad}}$ represent the states of radiation associated with Alice's particle CM amplitudes $|R(L)\rangle_A$, $|L(R)\rangle_B$ are the states of Bob's particle CM, and $|x=0\rangle_A$ is the final state of Alice's particle CM after the ``reversing'' Stern-Gerlach apparatus. Note that this state is akin to a GHZ state, a three-partite state which is (maximally) entangled but where no pair of subsystems is entangled.

However, in the actual case, where $T_A \gg {\mathcal D}_A$ or $T^2_A \gg {\mathcal Q}_A$, no photons/gravitons are emitted. How can we understand how Bob's particle became entangled with Alice's? It should be emphasized that Alice's and Bob's particles do not interact directly with each other; they each interact only with the electromagnetic/gravitational field. Since there is no radiation, it clearly must be the ``non-radiative part'' of the electromagnetic/gravitational field that is responsible for the ultimate entanglement of Alice's and Bob's particles. 
 
Entanglement of a particle with the non-radiative part of the field it generates has been previously considered by Unruh~\cite{unruh2000false} who referred to this as ``false loss of coherence.'' Unruh used the terminology ``false'' because, although the non-radiative part of the field formally produces a decoherence in a particle that would otherwise be in a coherent superposition, the coherence would be restored if the particle were recombined adiabatically. However, this assumes the absence of other matter that interacts with the field. In the case of our Gedankenexperiment with $T_A \gg {\mathcal D}_A$ or $T^2_A \gg {\mathcal Q}_A$, the non-radiative part of the field of Alice's particle interacts with Bob's particle. Even though Alice recombines her particle adiabatically, she cannot restore the coherence of her particle because the correlations of her particle with the non-radiative part of the field get transferred to Bob's particle. In Dirac notation
\begin{equation}
\underbrace{|x=L,\uparrow\rangle_A|\phi_L\rangle_{F} |R\rangle_B+|x=R,\downarrow\rangle_A|\phi_R\rangle_{F} |L\rangle_B}_{t<T_A}\rightarrow\underbrace{|x=0\rangle_{A}|\phi_0\rangle_{F}\otimes(|\uparrow\rangle_A |R\rangle_B+|\downarrow\rangle_A |L\rangle_B)}_{t>T_A},
\end{equation}
where here $|\phi_{L(R)}\rangle_{F}$ represent the states of the non-radiative part of the gravitational (or electromagnetic) field associated to Alice's center-of-mass amplitudes $|L(R)\rangle_{A}$, respectively.
This implies that the original loss of coherence of Alice's particle and the non-radiative part of its electromagnetic/gravitational field was {\em not} false!

The above discussion highlights the fact that it is extremely artificial to separate the electromagnetic/gravitational field into ``radiative'' and ``non-radiative'' parts. While Bob is performing his experiment, he has no way of knowing whether his particle is interacting with the radiative part of the field of Alice's particle (as would be the case if $T_A < {\mathcal D}_A$ or $T^2_A < {\mathcal Q}_A$) or with the non-radiative part of the field of Alice's particle (as would be the case if $T_A \gg {\mathcal D}_A$ or $T^2_A \gg {\mathcal Q}_A$). Both the radiative and non-radiative parts of the field are equally capable of entangling Bob's particle with Alice's. 

These considerations imply that a quantum massive particle in spatial superposition should be considered entangled with its own Newtonian-like gravitational field. One does not need freely propagating gravitons for the quantum gravitational field to carry quantum information. This conclusion is in agreement with the fact that entanglement cannot be increased by local operations and classical communication (LOCC). Releasing Bob's particle from the trap is a local operation and it cannot entangle it with Alice's far away particle unless we assume that A is already entangled with the field, and that releasing Bob's particle is just an entangling operation between it and the field. Thus, our conclusion supports the claim that recent proposals for table-top experiments aiming to entangling particles via their gravitational interaction can probe a quantum feature of gravity.

It is not clear to us what the full ramifications of these considerations are for the formulation of a quantum theory of gravity, but it is our hope and expectation that a deeper understanding of simple examples of the sort we have analyzed will help guide us in the right direction.


\section*{Acknowledgement}
AB is supported by H2020 through the MSCA IF pERFEcTO (GrantNo.  795782). CB, FG, EC and MA acknowledge the support of the Austrian Academy of Sciences through  Innovationsfonds  Forschung,  Wissenschaft  und  Gesellschaft,  and the University of Vienna through the research platform TURIS. FG and EC acknowledge support from the the doctoral program ``Complex Quantum Systems" (CoQuS). This project has received funding from the European Union’s Horizon 2020 research and innovation programme under grant agreement No 766900 (project TEQ) and from the European Research Council (ERC) under the European Union’s Horizon 2020 research and innovation programme (grant agreement No 649008). This publication was made possible through the support of a grant from the John Templeton Foundation. The opinions expressed in this publication are those of the authors and do not necessarily reflect the views of the John Templeton Foundation. The research of RMW was supported in part by NSF grants PHY 15-05124 and PHY18-04216 to the University of Chicago.


\bibliographystyle{apsrev4-1}
\bibliography{references2.bib}


\end{document}